\documentclass[twocolumn, pra, showpacs,superscriptaddress,floatfix]{revtex4}
\usepackage{mathrsfs}
\usepackage{amsmath}
\usepackage{graphicx}

\begin{document}

\title{Mixture of Tonks-Girardeau gas and Fermi gas in one-dimensional optical
lattices}
\author{Shu Chen}
\email{schen@aphy.iphy.ac.cn}
\affiliation{Institute of Physics, Chinese Academy of
Sciences, Beijing 100190, China}
\author{Junpeng Cao}
\affiliation{Institute of Physics, Chinese Academy of Sciences,
Beijing 100190, China}
\author{Shi-Jian Gu}
\email{sjgu@sun1.phy.cuhk.edu.hk}
\affiliation{Department of Physics and ITP, Chinese
University of Hong Kong, Hong Kong, China}

\begin{abstract}
We study the Bose-Fermi mixture with infinitely boson-boson
repulsion and finite boson-Fermion repulsion. By using a generalized
Jordan-Wigner transformation, we show that the system can be mapped
to a repulsive Hubbard model and thus can be solved exactly for the
case with equal boson and fermion masses. By using the Bethe-ansatz
solutions, we investigate the ground state properties of the mixture
system. Our results indicate that the system with commensurate
filling $n=1$ is a charge insulator but still a superfluid with
non-vanishing superfluid density. We also briefly discuss the case
with unequal masses for bosons and fermions.

\end{abstract}
 \maketitle

Mixtures of quantum degenerate atoms recently became a subject of intense studies of
both experiment and theory. One of particularly interesting systems is mixture of
ultracold bosonic and fermionic atoms \cite{Hulet,BF_exp,Fukuhara}, which have become
accessible through the development of sympathetic cooling \cite{Hulet,BF_exp}. The
experimental progress in manipulating cold atoms in effective one-dimensional (1D)
waveguides and the ability of tuning the effective 1D interactions by Feshbach
resonance leads experiment accessible to the strong correlation regime of 1D quantum
gas \cite{Paredes,Toshiya}. Meanwhile, by loading the atomic system into the optical
lattice \cite{Greiner,Jaksch}, one can simulate not only the solid state systems in a
highly tunable way but also new systems which may not be realized in condensed matter,
such as mixtures of Bose-Fermi atoms. These advances open a new channel to investigate
numerous phenomena of low-dimensional correlated lattice models which play important
roles in condensed matter physics.

To gain a deep insight of properties of the low-dimensional quantum
mixtures, some refined methods capable of dealing with strong
correlations are especially important. For example, the method of
Bose-Fermi mapping has been extensively exploited to study the
Tonks-Girardeau (TG) gas \cite{Girardeau,Rigol}. This method has
been also generalized to study the multi-component quantum gas in
the infinitely repulsive limit \cite{Girardeau07,Chen}. The extended
Bose-Fermi mapping method is only limited to a special case with no
tunable parameter of interaction, where all the intra- and
inter-component interactions go to infinite. In addition, the
Bose-Fermi mixture with equal boson-boson and boson-fermion
interactions can be exactly solved by Bethe-ansatz
\cite{Demler,Lai,GuanXW}. Unfortunately, its corresponding lattice
model is no longer integrable. So far, the ground state phase
diagram of the 1D Bose-Fermi Hubbard model in optical lattice has
been studied by mean-field theory \cite{Albus,Lewenstein},
Bosonization method \cite{Cazalilla,WangDW}, exact diagonalization
method \cite{Roth} and quantum Monte Carlo method
\cite{Troyer,Zujev,Senguptal}.
Despite the intensive studies of the lattice model
\cite{Albus,Lewenstein,Cazalilla,WangDW,Roth,Troyer,Zujev,Kuklov,Hebert},
no analytically exact result has been given except the TG limit
\cite{Girardeau07}, in which however the model suffers the problem
of a huge degeneracy of ground states (GSs). In this work, we shall
study the boson-fermion mixtures with the aim to give some exact
conclusions apart from the TG limit and focus on the case with an
infinite boson-boson repulsion but a tunable boson-fermion
interaction, which is found to be exactly solvable when the hopping
amplitude $t_b$ for boson equals to $t_f$ for fermion.

We consider a mixture system of bosonic and spin-polarized fermionic atoms confined in
a deep 1D optical lattice.  For sufficiently strong periodic potential and low
temperatures, the atoms will be confined to the lowest Bloch band and the low energy
Hamiltonian is described by the Hamiltonian
\begin{eqnarray}
H &=&-\sum_{i,\sigma =b,f}\left( t_\sigma a_{i\sigma }^{\dagger
}a_{i+1\sigma }+H.c.\right)   \nonumber \\
&&+\frac 12\sum_{i}U_bn_{i,b}\left( n_{i,b}-1\right)
+U_{bf}\sum_in_{i,b}n_{i,f},  \label{2cBHM}
\end{eqnarray}
where $a_{i\sigma }$ are bosonic or fermionic annihilation operators
localized on site $i$, and $n_{i\sigma }=a_{i_\sigma }^{\dagger
}a_{i_\sigma }$. In principle, the interaction parameters $U_{bf}$
and $U_b$ can be tuned experimentally by the Feshbach resonance. In
this work, we shall focus on the case with infinitely strong
boson-boson repulsion, i.e., $U_b=\infty ,$ and a tunable
inter-species repulsion $U_{bf}=U.$ In this case, the boson is a
hard-core one or a TG gas, for which the states occupied by more
than one boson are prohibited. Similarly, the states occupied by
more than one fermion are not permitted due to the Pauli principle.
However, a boson and a fermion can occupy the same site which
contributes an on-site energy $U$. In the hard-core limit, the
Bose-Fermi mixture model can be simplified to
\begin{equation}
H_{BF}=-\sum_{i,\sigma =b,f}\left( t_{\sigma} a_{i\sigma }^{\dagger
}a_{i+1\sigma }+H.c.\right) +U\sum_in_{i,b}n_{i,f},  \label{HHB}
\end{equation}
with additional on-site constraints $a_{ib}^{\dagger }a_{ib}^{\dagger
}=a_{ib}a_{ib}=0$ and $\left\{ a_{ib},a_{ib}^{\dagger }\right\} =1$ assigned
to avoid double or higher occupancy.

{\it Mixture with equal masses.---} Firstly, we focus on the case
that the bosonic and fermionc atoms have the same masses,
which is approximately satisfied for the Bose-Fermi mixture of heavy isotopic atoms,
for example, the $^{174}$Yb-$^{173}$Yb mixture \cite{Fukuhara}. The Bose-Fermi mixture
with equal masses provides a solvable limit, which may serve as a touchstone for
various numerical simulations. For the model system with the fermion and the boson
having the same mass, we have $t_b=t_f$. It is convenient to use the following extended
Jordan-Wigner (JW) transformations
\begin{equation}
a_{ib} =\prod_{j<i}e^{i\pi c_{j\uparrow }^{\dagger }c_{j\uparrow
}}c_{i\uparrow },\quad  a_{if} = \prod_{j=1}^Ne^{i\pi c_{j\uparrow
}^{\dagger }c_{j\uparrow }}c_{i\downarrow }, \label{jw2}
\end{equation}
which maps the Hamiltonian of hard-core Bose and Fermi mixture model into a
Hubbard model
\begin{equation}
H_F=-\sum_{i,\sigma }\left( t_\sigma c_{i\sigma }^{\dagger }c_{i+1\sigma
}+H.c.\right) +U\sum_in_{i\uparrow }n_{i\downarrow }.  \label{Hubbard}
\end{equation}
The second mapping in eq. (\ref{jw2}) is introduced to enforce the
Fermion operators $c_{i,\uparrow}$ and $c_{i,\downarrow}$ fulfil the
anti-commutation relation $\left\{
c_{i,\uparrow},c_{i,\downarrow}\right\} =0$. The Hamitonians of
$H_{BF}$ and $H_F$ have the same spectrum of energy. Therefore, we
can get the eigen-energy of $H_{BF}$ with $t_b=t_f=t$ from the
well-known Lieb-Wu solution of the Hubbard model \cite{LiebWu},
i.e., the eigenenergy is given by $ E=-2t\sum_j^N\cos k_j$ with
$k_j$ determined by the Bethe-ansatz equations
\begin{eqnarray}
2\pi I_{j}&=& k_{j}L - \sum_{\beta =1}^{M}\theta _{1}(\Lambda
_{\beta }-\sin
k_{j}),  \label{eq:BAE1} \\
2\pi J_{\alpha }&=& \sum_{j=1}^{N}\theta _{1}(\Lambda _{\alpha
}-\sin k_{j})-\sum_{\beta =1}^{M}\theta _{2}(\Lambda _{\alpha
}-\Lambda _{\beta }), \label{eq:BAE2}
\end{eqnarray}%
where $\theta _{n}(k)=2\tan ^{-1}(4k/nU)$, $\quad j=1,\cdots, N$,
$\beta=1,\cdots, M$, $N=N_{b}+N_{f}$, $M=N_f$, $N_b$ ($N_f$) is the
number of bosons (fermions), and $L$ is the size of the system. Here
the set $\{I_{j},J_{a}\}$ play the role of quantum number. We also
solve the ground state energy of $H_{BF}$ by using the numerical
exact diagonalization method and compare with the result obtained by
solving the Bethe-ansatz equations. It is found that the numerical
result agrees with the Bethe-ansatz solution exactly.

The unitary mapping builds a bridge between the hard-core Bose-Fermi
mixture and the extensively studied Hubbard model. Since these two
models sharing the same energy spectrum, we can conclude that they
have the same thermodynamic properties. The GS properties of
hard-core Bose-Fermi model also share some similarities with the
Hubbard model, i.e., there exists no Mott transition from superfluid
to Mott insulator for any finite U. For the incommensurate filling
case, the system is in a superfluid phase, whereas the system with
the commensurate filling $n=1$ is in a Mott phase for any finite U,
which is characterized by the presence of a charge gap and
simultaneously a gapless mode of mixture composition fluctuations.

The superfluid density of the bosonic component can also be characterized by the
bosonic phase stiffness, which reflects the response of a superfluid component to the
imposed phase gradient and is defined as \cite{Shastry}
\[
D_{b} = \left. {\frac{L}{2}
\frac{\partial^2 {E_0(\phi_b)}}{
\partial \phi_b^2}}\right|_{\phi_b=0}
\]
which is proportional to the Drude weight. Similarly, the fermionic stiffness can be
represented as
\[
D_{f} = \left. {\frac{L}{2} \frac{\partial^2 {E_0(\phi_f)}}{
\partial \phi_f^2}}\right|_{\phi_f=0}.
\]
Here, $E_0$ is the ground-state energy, $\phi_b$ and $\phi_f$ are the
component-dependent flux in units of $\hbar c/e$ for the boson and fermion
respectively, which can be incorporated in the wavefunction by making the usual gauge
transformation $ a_{i,\sigma} \rightarrow e^{i\phi_\sigma r_i /L}a_{i,\sigma}$. A
finite bosonic or fermionic stiffness is characteristic of a superfluid or conductor,
whereas the stiffness vanishes for a insulator. In the presence of component-dependent
flux, the eigenenergy of the system is also given by $E=-2t\sum_j \cos k_j$ with
$k_{j}$ determined by the revised Bethe-ansatz equations \cite{Shastry}
\begin{eqnarray}
2\pi I_{j}&=& k_{j}L-\phi_b-\sum_{\beta =1}^{M}\theta _{1}(\Lambda
_{\beta }-\sin
k_{j}),  \label{eq:BAE1} \\
2\pi J_{\alpha }&=& \sum_{j=1}^{N}\theta _{1}(\Lambda _{\alpha
}-\sin k_{j})-\sum_{\beta =1}^{M}\theta _{2}(\Lambda _{\alpha
}-\Lambda _{\beta }) \nonumber \\
&&+\phi_f-\phi_b . \label{eq:BAE2}
\end{eqnarray}%

\begin{figure}[tbp]
\includegraphics[width=9cm]{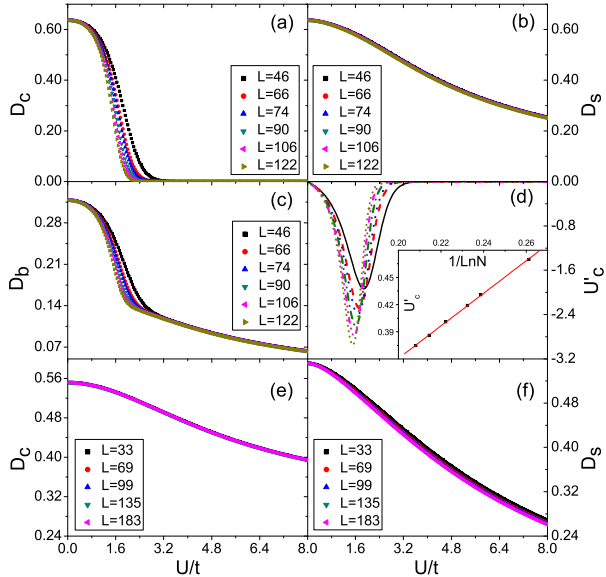}
\caption{ (color online) The stiffness of system. The panels (a),
(b), (c) and (d) correspond to the filling factor $n=1$, whereas the
panels (e) and (f) correspond to $n= 2/3$.
Finite size analysis of the minimum of the derivative of the charge stiffness is shown
in inset of (d). Here we have taken $N_b=N_f$.} \label{fig_stiffness}
\end{figure}



By solving the revised BAEs, we can directly calculate the stiffness of system. In
order to calculate the charge stiffness $D_{c}$, we set the magnetic flux of hard-core
bosons $\phi_b$ and that of spinless fermions $\phi_f$ to be the same, i.e.,
$\phi_b=\phi_f=\phi$, whereas the spin stiffness $D_{s}$ is calculated by taking
$\phi_b=-\phi_f=\phi$. Explicitly, we have
\[
D_{c} = \left. {\frac{L}{2}
\frac{\partial^2 {E_0(\phi)}}{
\partial \phi^2}}\right|_{\phi=\phi_b=\phi_f=0}
\]
and
\[
D_{s} = \left. {\frac{L}{2} \frac{\partial^2 {E_0(\phi)}}{
\partial \phi^2}}\right|_{\phi=\phi_b=-\phi_f=0}.
\]
For the case with commensurate filling, we display the charge, spin and boson stiffness
in (a), (b) and (c) of Fig. \ref{fig_stiffness}, respectively.  It is obvious that the
charge stiffness goes to zero quickly when the interaction U exceeds a critical value.
To extrapolate the critical $U_c$ in the thermodynamic limit, we make finite size
analysis of the transition point where the charge stiffness tends to vanish, which is
characterized by the minimum of the derivative of the charge stiffness as shown in the
Fig. \ref{fig_stiffness}(d). The scaling behaviors of transition points can be fitted
as $U_{cm}=0.00104+1.79789/\ln N$. When the system-size tends to infinity, the critical
on-site interactions reads $U_c=0.00104\pm 0.00354$, which covers the zero within the
scope of fitting error. This means that the system is in a Mott phase with zero charge
stiffness for any finite repulsion between the hard-core boson and fermion in the
thermodynamic limit. However, as shown in Fig. \ref{fig_stiffness}(b) and
\ref{fig_stiffness}(c), the spin and boson stiffness do not vanish even the system is
in a Mott phase where the fluctuation of particle number (charge) on each site is
greatly suppressed. Furthermore, our results show that the density of superfluid
fulfills an interesting relation with the charge and spin stiffness, $D_b=(D_c+D_s)/4$
(see appendix). The non-vanishing $D_b$ indicates that the boson is still in a
superfluid phase even in the phase of charge insulator. This is induced by the
fluctuation of bosons. Despite the charge fluctuation is suppressed, the boson can
tunnel to the neighboring sites by a virtual second order process. For the
incommensurate case with $n<1$, both the charge and spin stiffness do not vanish as
shown in Fig. \ref{fig_stiffness}(e) and \ref{fig_stiffness}(f), and no a Mott phase
exists.

Particular attention should be paid to the special cases with a
partial commensurate filling, for example, cases with $n_f=1$,
$n_b<1$.
Then the system favors that every site has one fermion which
supplies a background and the bosons can hop on it freely. The
energy of this state is
\[
E=N_b U-\frac{tL}{\pi}\int_{-k_b}^{k_b} \cos(k +\phi_b /L)dk,
\]
where $k_b=\pi n_b$. It is straightforward to get $
D_{b}={t}\sin(\pi n_b)/{\pi}$ and $D_{f}=0$, which indicates that
the bosons form a superfluid whereas the fermions form an insulator.
Similarly, for the case with $n_b=1$ and $n_f<1$, the bosons are in
an insulator state which provides a homogeneous background for the
fermions which form a conductor with $ D_{f}={t}\sin(\pi
n_f)/{\pi}$.

\begin{figure}[tbp]
\includegraphics[width=9.0cm]{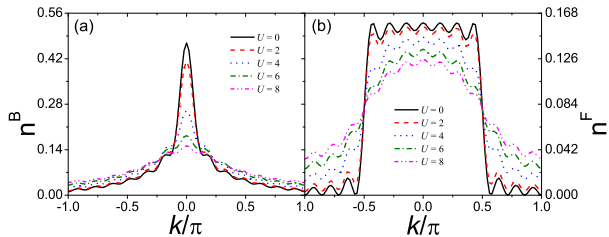}
\caption{ (color online)Momentum distributions $n^{B}(k)$ for
hard-core bosons (a) and $n^{F}(k)$ for spinless fermions (b). Here
$L=14$, $N_{\rm b}=7$ and $N_{\rm f}=7 $.} \label{momentum}
\end{figure}

Despite $H_{BF}$ and $H_F$ sharing the same energy level structure,
they have different ground state wavefunction due to the
intrinsically different exchange symmetry of the wave functions for
Bose and Fermi systems. Supposed the wavefunction of the Fermi
Hubbard model is given by $\Psi _F(x_1,\cdots ,x_n;x_{n+1},\cdots
,x_N),$ the wavefunction of $H_{HB}$ can be constructed as
\begin{eqnarray*}
&&\Psi _{BF} (x_1,\cdots ,x_n;y_{n+1},\cdots ,y_N) \\
&=&\prod_{i<j}sgn(x_i-x_j) \Psi _F(x_1,\cdots ,x_n;y_{n+1},\cdots
,y_N),
\end{eqnarray*}
where $sgn(x_i-x_j)=(x_i-x_j)/|x_i-x_j|$ is the sign function.
Consequently, the observable associated with the wave functions
rather than the energy level structures should display different
behaviors, which can be displayed in the off-diagonal density matrix
and the momentum distributions of the hard-core boson. Explicitly,
the density matrices of boson and fermion are defined as $ \rho
_{ij}^{B}=\left\langle a_{i,b}^{\dagger }a_{j,b}\right\rangle$ and
$\rho _{ij}^{F}=\left\langle a_{i,f}^{\dagger }a_{j,f}\right\rangle$
respectively, which exhibit quite different behaviors for boson and
fermion. The momentum distribution can be obtained by the Fourier
transformation of the corresponding density matrix. For example, we
have
\begin{equation}
n^{B,F}(k)=\frac{1}{2\pi L}\sum_{i,j} \rho _{ij}^{B,F} e^{-ik(i-j)}.
\end{equation}
In Fig. (\ref{momentum}), we display the momentum distribution of
the Bose and Fermi systems respectively for the case with
commensurate filling $n_b=n_f=1/2$. It is shown that the momentum
distribution of the hard-core bosons has a sharp peak, which
reflects the bosonic nature of the particles. With the increase in
inter-component interactions, the momentum distribution spreads
wider and wider but the pronounced peak around the zero momentum is
kept. While the momentum distribution of the fermion exhibits the
free Fermi distribution for $U=0$, it becomes more wider and
develops a wide tail with the increase in $U$.

{\it Mixture with unequal masses.---} Finally, we give a brief
discussion on the case with $t_b \neq t_f$ which corresponds to the
system where the bosonic and fermionc atoms have different masses,
such as mixture of $^7$Li and $^{40}$K. In general, the single
particle hopping amplitude is inversely proportional to the mass of
atoms, {\it i.e.}, $t_f / t_b = m_b / m_f$. Taking the mixture
$^7$Li and $^{40}$K as an example, we have $t_{\beta}= t_f / t_b
\approx 0.175$. For the general case with $t_{\beta} \neq 1$, the
mixture of hard-core bosons and fermions is no longer exactly
solvable, although a mapping to the asymmetric Hubbard model
\cite{Guprb} still holds true via the generalized JW transformation.

In the heavy Fermi mass limit $t_{\beta} \rightarrow 0$, the
fermions lose the mobility and the asymmetric mixture model is
related to the Falicov-Kimball model \cite{Falicov} via the
generalized JW transformation. Away from integer-filling with $n<1$,
the heavy fermion and the light bosons are favorable to stay in
different regimes to lower the kinetic energy of bosons. Therefore
the system is expected to have a transition from density wave to
phase separation. Next we shall use the density matrix
renormalization group method to obtain a quantitative phase diagram
for the mixture of TG gas and Fermi gas with $N_b = N_f$ and a
filling $n=4/5$.

\begin{figure}[tbp]
\includegraphics[width=8.8cm]{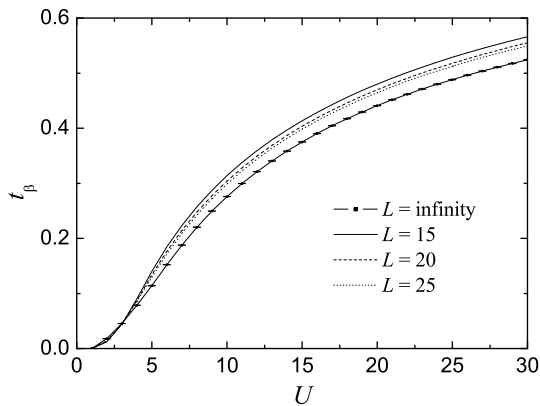}
\caption{Phase diagram for the asymmetric Bose-Fermi mixture with
$n=4/5$. } \label{phase}
\end{figure}

Taking into account that the dominating configuration of fermionic
atoms is quite different in the density-wave phase or phase with
phase separation, we introduce the following structure factor of
density wave (DW) of fermionic atoms
\begin{eqnarray}
S_{\rm FDW}(q) =\frac{1}{L}\sum_{jl}\left[ e^{iq(j-l)}(\langle
n_{j,f} n_{l,f} \rangle - \langle n_{f}\rangle^2) \right],
\end{eqnarray}
where $q=2n\pi/L$ and $n=0, 1, \cdots, L$. We calculate the
structure factor as a function of $t_\beta$ for different modes for
systems with different sizes $L$ and strengthes of interactions $U$.
The results show an obvious competition between the modes of $S_{\rm
FDW}(q=2\pi/L)$ and $S_{\rm FDW}(q=N\pi/L)$. In the heavy fermionic
atom limit with small $t_\beta$, $S_{\rm FDW}(q=2\pi/L)$ dominates,
which indicates phase separation in this region \cite{PLemberger92}
where configurations of fermionic atoms like $|f, f, f, f, \circ,
\circ, \circ, \circ, \circ, \circ\rangle $  are found to be
dominant. On the other hand, as $t_\beta\rightarrow 1$, $S_{\rm
FDW}(q=N\pi/L)$ exceeds $S_{\rm FDW}(q=2\pi/L)$, which implies that
fermionic atoms distribute uniformly on the optical lattice. Then
together with bosonic atoms, the ground state becomes the so called
DW state similar to the state of symmetric model, which is the
limiting case with $t_\beta=1$ studied in the above section.
Consequently, we can determine the transition point on the
$U-t_\beta$ plane for a finite system from the intersection of the
structure factor of two modes. In Fig. \ref{phase}, we plot the
phase diagram on the $U-t_\beta$ plane for systems with a filling
factor $n=4/5$ for different system-sizes $L=15, 20, 25$. The
infinite size limit is obtained by extrapolation from the
finite-size analysis. Below the phase boundary line, the
phase-separation phase dominates, while the DW phase is dominated
above the boundary line.

In summary, we study the mixture of TG gas and fermions in a 1D
optical lattice. We show that this system can be mapped to the Fermi
Hubbard model by a generalized Jordan-Wigner transformation and thus
is exactly solvable when the bosons and fermions have the same
masses. Based on the Bethe-ansatz solution, we calculate the charge
stiffness and the superfluid density for the systems with either
commensurate or incommensurate filling. Our results show that the
system with commensurate filling factor $n=1$ is a charge insulator
but remains to be a superfluid characterized by a non-vanishing
superfluid density. We also give a brief discussion to the case with
unequal boson and fermion masses. In the heavy Fermi mass limit, our
result indicates that a  phase-separation phase arise. The phase
transition from the density wave phase to the phase-separation phase
is discussed.

This work is supported by NSF of China under Grants Nos. 10974234,
10821403, 10974233 and 10934010, National Program for Basic Research
of MOST China under Grants No.2010CB922904, and the Earmarked Grant
for Research from the Research Grants Council of HKSAR, China
(Project No. CUHK 401108).

\appendix
\section{}
In this appendix, we provide a derivation of the relation $D_b=\frac{1}{4}(D_c+D_s)$.
In the presence of flux, the Hamiltonian (2) becomes
\begin{eqnarray*}
H_{BF}&=&-t\sum_{i}\left( a_{ib }^{\dagger }a_{i+1b
}e^{i\phi_b/L}+H.c.\right)\\
& & -t\sum_{i}\left( a_{i f }^{\dagger }a_{i+1 f}e^{i\phi_f/L}+H.c.\right) +U
\sum_in_{i,b}n_{i,f}.
\end{eqnarray*}
Expanding the Hamiltonian in terms of $\phi_b/L$ and $\phi_f/L$, we have
\begin{eqnarray*}
H_{BF}&=&\left(T_b-\frac{\phi_b j_b}{L}-\frac{T_b }{2}\frac{\phi_b^2}{L^2}\right) +
\left(T_f-\frac{\phi_f j_f}{L}-\frac{T_f
}{2}\frac{\phi_f^2}{L^2}\right)\\
& & +U\sum_in_{i,b}n_{i,f} + O(\phi_b^4,\phi_f^4),
\end{eqnarray*}
where $T_{\sigma}=-t\sum_{i}\left( a_{i \sigma }^{\dagger }a_{i+1 \sigma }+H.c.\right)$
and $j_{\sigma}=i t \sum_{i}\left( a_{i \sigma }^{\dagger }a_{i+1 \sigma }-
H.c.\right)$ with $\sigma=b,f$. So according to the perturbation theory,
\[
D_c = \frac{1}{L}\left[\frac{1}{2}\langle-T_c\rangle -\sum_{n\neq 0}\frac{\langle 0|j_c|n\rangle^2}{E_n-E_0}\right],\\
\]
\[
D_s = \frac{1}{L}\left[\frac{1}{2}\langle-T_s\rangle -\sum_{n\neq 0}\frac{\langle 0|j_s|n\rangle^2}{E_n-E_0}\right],\\
\]
\[
D_b = \frac{1}{L}\left[\frac{1}{2}\langle-T_b\rangle -\sum_{n\neq 0}\frac{\langle 0|j_b|n\rangle^2}{E_n-E_0}\right],\\
\]
\[
D_f = \frac{1}{L}\left[\frac{1}{2}\langle-T_f\rangle -\sum_{n\neq 0}\frac{\langle
0|j_f|n\rangle^2}{E_n-E_0}\right],
\]
where the current operators: $j_c=j_b+j_f$, $j_s=j_b-j_f$, and $T_c=T_s=T_f+T_b=2T_b$.
Then
\begin{eqnarray*}
D_c&+&D_s\\
&=&\frac{1}{L}\left[\frac{1}{2}\langle-4T_b\rangle-2\sum_{n\neq 0}\frac{\langle 0|j_c|n\rangle^2}
{E_n-E_0}-2\sum_{n\neq 0}\frac{\langle 0|j_s|n\rangle^2}{E_n-E_0} \right]\\
&=&\frac{1}{L}\left[\frac{1}{2}\langle-4T_b\rangle-4\sum_{n\neq 0}\frac{\langle
0|j_b|n\rangle^2}{E_n-E_0} \right] =4D_b .
\end{eqnarray*}
Finally, we have
\[
D_b=\frac{1}{4}(D_c+D_s) .
\]


\end{document}